\documentclass[%
 reprint,
 showkeys,
 superscriptaddress,
 amsmath,amssymb,
 aps,
 prd,
 floatfix,
]{revtex4-2}

\usepackage{graphicx}% Include figure files
\usepackage{dcolumn}% Align table columns on decimal point
\usepackage{bm}% bold math
\usepackage{placeins}
\usepackage{hyperref}
\hypersetup{
    colorlinks = true,   
    citecolor  = blue,   
    linkcolor  = blue,   
    urlcolor   = blue    
}
\newcommand{\ssr}{Space Sci. Rev.}
\usepackage{xcolor}

\begin{document}

\title{High-energy Electron Transport with Discrete Inverse Compton Scatterings}

\author{Ruihao Jing}
\email{jingruihao@my.swjtu.edu.cn}
\author{Siming Liu}
\email{liusm@swjtu.edu.cn}

\affiliation{School of Physical Science and Technology, Southwest Jiaotong University, Chengdu 610031, People's Republic of China}

\date{\today}

\begin{abstract}
After escaping from pulsar wind nebulae (PWNe), the transport of high-energy electrons in the interstellar medium (ISM) is usually described with the diffusion equation with a continuous radiative energy loss. %Conventional models generally approximate inverse Compton scattering (ICS) as a continuous energy-loss process. 
However, when the electron energy approaches $\sim 100~\text{TeV}$ and the radiative energy loss is dominated by the inverse Compton scattering (ICS), the discrete nature of ICS becomes prominent, and the approximation of continuous radiative energy loss is invalid. %the energy of evolved electrons should be considered as a broad distribution rather than a deterministic value. 
Starting from the Klein--Nishina differential scattering rate, this paper constructs an integral-differential equation to describe the stochastic discrete nature of the inverse Compton cooling, %replacing the continuous inverse Compton energy-loss term in the traditional transport equation with this discrete inverse Compton collision operator. 
and the equation is solved with the Fourier and Laplace transform method. Application to the TeV halo of Geminga pulsar leads to a diffusion coefficient  {$\sim 15$\%} lower than that of the continuous energy loss model. 

\end{abstract}

\maketitle %

\section{Introduction} \label{sec:intro}

Pulsar halos are extended $\gamma$-ray sources produced via ICS by high-energy electrons and positrons that escape from PWNe and propagate through the ISM (for reviews, see Refs.~\citep{10.3389/fspas.2022.1022100, doi:10.1142/S0217751X22300113, Lopez-Coto2022}). Once particles escape from the PWN, their energy evolution is primarily governed by radiative loss processes such as synchrotron radiation and ICS. In particular, ICS not only determines the $\gamma$-ray emission characteristics of pulsar halos, but also plays a pivotal role in the energy evolution of the high-energy parent electrons responsible for generating these $\gamma$-rays.

Within the parameter space typically considered, electron synchrotron radiation in magnetic fields can be well approximated as a continuous cooling process; in contrast, ICS between electrons and background photons fundamentally consists of a series of discrete, stochastic scattering events. In practical calculations, ICS is also frequently treated under the continuous cooling approximation \citep{Schlickeiser_2010, 1612614924733-1139738958}. However, as the electron energy increases and enters the Klein–Nishina regime, the scattering cross section is suppressed and the scattering rate decreases, whereas the fraction of energy transferred in a single collision grows substantially. For example, an electron with an energy of $100\,\mathrm{TeV}$ scattering off a cosmic microwave background (CMB) photon can lose nearly half of its initial energy within the kinematically allowed limit \citep{PhysRevD.107.103021}, an issue already noted in classic studies on astrophysical radiation processes \citep{RevModPhys.42.237}. Under these conditions, the electron energy no longer evolves continuously along a single trajectory dictated by the average cooling rate; instead of a deterministic quantity, the evolved electron energy develops a distribution with a non-negligible width. This broadening effect not only alters the energy spectrum and spatial propagation of ultra-high-energy electrons, but also directly impacts the calculated $\gamma$-ray spectrum.

Precise measurements of the cosmic-ray positron flux and positron fraction by experiments such as PAMELA and AMS-02 have revealed a prominent high-energy positron excess \citep{2009Natur.458..607A,PhysRevLett.110.141102,PhysRevLett.122.041102}; meanwhile, high-precision measurements of the electron flux or the total electron-plus-positron flux by experiments including AMS-02, DAMPE, CALET, and H.E.S.S. have provided critical observational constraints on the origin and propagation of high-energy electrons and positrons \citep{PhysRevLett.122.101101, 2017Natur.552...63D, PhysRevLett.131.191001, PhysRevLett.133.221001}. Nearby pulsars such as Geminga are widely considered potential sources of high-energy electrons and positrons near Earth and could contribute to the observed positron excess \citep{Fang_2018,8654529}. Previous studies have shown that the discrete nature of ICS can broaden the propagated particle energy distribution and modify the spectral structure in the high-energy regime \citep{PhysRevD.107.103021, g41m-jzfj}. Therefore, if discrete scattering preserves a more pronounced high-energy tail in the evolved particle spectrum, the continuous cooling approximation may underestimate the contribution of nearby sources such as Geminga to the high-energy electron and positron fluxes near Earth, as well as alter the overall spectral shape of the predicted flux. Thus, discrete ICS may affect the theoretical predictions of electron and positron energy spectra as well as the quantitative interpretation of the positron excess. Motivated by this rationale, this work incorporates discrete ICS corrections into the transport equation.

Starting from the Klein–Nishina differential scattering rate, this study constructs a discrete ICS collision operator for electrons to replace the continuous ICS cooling term in the conventional transport equation, and presents the corresponding numerical solution method. The remainder of this paper is organized as follows: Section~II establishes the electron transport equation incorporating the discrete ICS collision operator; Section~III presents the numerical method for solving this equation; Section~IV validates the reliability of the proposed model and numerical scheme through comparison with Monte Carlo simulations~\citep{g41m-jzfj}, while comparing the differences between the discrete scattering treatment and the continuous cooling approximation in terms of electron energy evolution and spatial diffusion; and Section~V summarizes the main conclusions of this work.

\section{The Electron Transport Equation with Discrete Inverse Compton Scatterings}

The scattering rate for ICS is given by \citep{RevModPhys.42.237}
\begin{equation}\label{eq:ics_kernel}
\begin{split}
& f(E_e, E_\gamma, \varepsilon) =  \frac{3\sigma_{\rm T}c}{4(E_e/m_e c^2)^2} \times \frac{n(\varepsilon)}{\varepsilon}\\
& \times \left[ 2\chi \ln \chi + (1 + 2\chi)(1 - \chi) + \frac{(\Gamma \chi)^2 (1 - \chi)}{2(1 + \Gamma \chi)} \right],
\end{split}
\end{equation}
Here, $f(E_e, E_\gamma, \varepsilon) \equiv {\rm d}N / ({\rm d}t \, {\rm d}E_\gamma \, {\rm d}\varepsilon)$ denotes the differential rate at which an electron of energy $E_e$ scatters with a background photon of energy $\varepsilon$ and produces a photon of energy $E_\gamma$. The dimensionless variables $\Gamma$ and $\chi$ are defined as
\begin{equation}\label{eq:gamma_q}
\Gamma = \frac{4E_e \varepsilon}{(m_e c^2)^2}, \quad \chi = \frac{E_\gamma / E_e}{(1 - E_\gamma / E_e)\Gamma},
\end{equation}
and $n(\varepsilon)$ denotes the background photon number density per unit energy interval, while $\sigma_{\rm T}$ denotes the Thomson cross section.

Denoting the net energy loss of the electron as $\omega$, we have $E_\gamma = \omega + \varepsilon$. Consequently, $f(E_e, E_\gamma, \varepsilon)$ can be expressed as $f(E_e, \omega + \varepsilon, \varepsilon)$, and integrating over $\varepsilon$ yields
\begin{equation}\label{eq:rate_integral}
W(\omega, E_e) = \int_0^\infty f(E_e, \omega + \varepsilon, \varepsilon) \, {\rm d}\varepsilon,
\end{equation}
The kinematically allowed range for ICS is given by
\begin{equation}\label{eq:kinematic_range}
\frac{1}{4\gamma^2} \le \chi \le 1, \qquad \gamma = \frac{E_e }{m_e c^2},
\end{equation}
and $f$ is defined to be zero for parameter combinations that do not satisfy these kinematic conditions. The quantity $W(\omega, E_e)$ represents the transition rate per unit time and per unit energy loss for a given initial electron energy $E_e$.

For a given initial electron energy $E_e$, integrating $W(\omega, E_e)$ over the entire range of energy loss yields the total ICS rate of the electron, which is given by
\begin{equation}\label{eq:collision_frequency}
\nu(E_e) = \int_0^{E_e - m_e c^2} W(\omega, E_e) \, {\rm d}\omega.
\end{equation}

Let $N(E, t)$ denote the number of electrons per unit energy interval at time $t$, i.e., the electron energy spectrum. For a sufficiently small time interval ${\rm d}t$, retaining terms up to first order in ${\rm d}t$ yields
\begin{equation}\label{eq:electron_evolution_dt}
\begin{aligned}
& N(E, t + {\rm d}t) =  [1 - \nu(E){\rm d}t] N(E, t) \\
& + {\rm d}t \int_0^{E_{\rm max} - E} W(\omega, E + \omega) N(E + \omega, t) \, {\rm d}\omega + o({\rm d}t),
\end{aligned}
\end{equation}
Here, the first term represents the contribution from electrons with energy $E$ at time $t$ that do not undergo ICS within the time interval ${\rm d}t$ and thus remain at energy $E$; the second term is the scattering gain term, representing the contribution from electrons with an initial energy $E + \omega$ that transition to energy $E$ after losing energy $\omega$ in a single scattering event; and $o({\rm d}t)$ denotes higher-order terms relative to ${\rm d}t$, which include contributions from two or more scatterings during this time interval as well as other unmentioned higher-order corrections.

Subtracting $N(E, t)$ from both sides of the short-time evolution relation and dividing by ${\rm d}t$, given that $o({\rm d}t) / {\rm d}t \to 0$, yields the continuous-time evolution equation for the electron energy spectrum:
\begin{equation}\label{eq:continuous_evolution}
\begin{aligned}
{\partial N(E, t) \over \partial t} = {} & \int_0^{E_{\rm max} - E} W(\omega, E + \omega) N(E + \omega, t) \, {\rm d}\omega \ \\
&- N(E, t) \nu(E),
\end{aligned}
\end{equation}
Here, the first term represents the transition of electrons with an initial energy $E + \omega$ to energy $E$ after losing energy $\omega$ in a single scattering event, while the second term accounts for electrons with an initial energy $E$ leaving this energy state due to scattering. Denoting the initial electron energy as $E' = E + \omega$ and defining $T(E, E') \equiv W(\omega, E + \omega)$, where $T(E, E')$ represents the transition rate per unit energy per unit time for an electron to transition from an initial energy $E'$ to a final energy state $E$. Since ${\rm d}E' = {\rm d}\omega$, the range of integration changes from $\omega \in [0, E_{\rm max} - E]$ to $E' \in [E, E_{\rm max}]$, yielding
\begin{equation}\label{eq:discrete_ics_master}
{\partial N(E, t) \over \partial t} = \int_E^{E_{\rm max}} T(E, E') N(E', t) \, {\rm d}E' \\
- \nu(E) N(E, t),
\end{equation}

Replacing the continuous ICS energy-loss term in the conventional transport equation with the discrete collision operator defined in Eq.~\eqref{eq:discrete_ics_master} yields
\begin{equation}\label{eq:anisotropic_ics_transport}
\begin{aligned}
{} &\frac{\partial N(\boldsymbol{r}, E, t)}{\partial t} \\
&=  D_\parallel(E) \frac{\partial^2 N(\boldsymbol{r}, E, t)}{\partial r_\parallel^2} + D_\perp(E) \nabla_\perp^2 N(\boldsymbol{r}, E, t) \\
& - \nu(E) N(\boldsymbol{r}, E, t) + \int_E^{E_{\rm max}} T(E, E') N(\boldsymbol{r}, E', t) \, {\rm d}E' \\
& + \frac{\partial}{\partial E} \left[ b(E) N(\boldsymbol{r}, E, t) \right] +Q(\boldsymbol{r}, E, t) ,
\end{aligned}
\end{equation}
Eq.~\eqref{eq:anisotropic_ics_transport} is the transport equation that describes the stochastic and discrete characteristics of ICS cooling. Here, $N$ is the electron number density, $E$ is the electron energy, and $r_\parallel$ and $r_\perp$ are the coordinates parallel and perpendicular to the mean magnetic field, respectively, with the origin set at the position of the injection source. $t$ is the time coordinate with its origin at the source birth time. The relation between the perpendicular and parallel diffusion coefficients relative to the mean magnetic field is $D_\perp = D_\parallel M_{\rm A}^4$ \citep{2008ApJ...673..942Y}, where $M_{\rm A}$ is the Alfv\'{e}nic Mach number. The term $\frac{\partial}{\partial E} \left[ b(E) N(\boldsymbol{r}, E, t) \right]$ represents the synchrotron radiation term, which is retained in the form of continuous energy loss. The synchrotron energy loss rate is given by $b(E) = b_0 E^2$. Here, $Q(\boldsymbol{r}, E, t)$ is the injection source term.

\section{Fourier–Laplace Transform Method}

To solve Eq.~\eqref{eq:anisotropic_ics_transport}, we first apply a spatial Fourier transform to the equation, i.e., $\widehat{N}(\boldsymbol{k}, E, t) = \int d^3r \, e^{-i\boldsymbol{k}\cdot\boldsymbol{r}} N(\boldsymbol{r}, E, t)$, where, $\boldsymbol{k}$ denotes the Fourier wave vector conjugate to the spatial coordinate $\boldsymbol{r}$. Under this Fourier transformation, the spatial differential operators satisfy, $\mathcal{F}_{\boldsymbol{r}} \left[ \frac{\partial^2 N}{\partial r_\parallel^2} \right] = -k_\parallel^2 \widehat{N} , \quad \mathcal{F}_{\boldsymbol{r}} \left[ \nabla_\perp^2 N \right] = -k_\perp^2 \widehat{N}$ , where the components $k_\parallel$ and $k_\perp$ are defined relative to the direction of the mean magnetic field. yielding
\begin{equation}\label{eq:10}
\begin{aligned}
[]&\frac{\partial \widehat{N}(\boldsymbol{k}, E, t)}{\partial t} \\
&=\widehat{Q}(\boldsymbol{k}, E, t) + \frac{\partial}{\partial E} \left[ b(E) \widehat{N}(\boldsymbol{k}, E, t) \right]\\
&  + \int_E^{E_{\rm max}} T(E, E') \widehat{N}(\boldsymbol{k}, E', t) \, dE'\\
&-\left[ \nu(E) + D_\parallel(E) k_\parallel^2 + D_\perp(E) k_\perp^2 \right] \widehat{N}(\boldsymbol{k}, E, t).  \\
\end{aligned}
\end{equation}

For a point source located at the spatial origin, the injection source term can be expressed as $Q(\boldsymbol{r}, E, t) = Q(E, t)\delta^{(3)}(\boldsymbol{r})$, which yields the injection term in Fourier space as $\widehat{Q}(\boldsymbol{k}, E, t) = Q(E, t)$. Here, $Q(E, t)$ represents the time-evolving injected electron energy spectrum.

After completing the spatial Fourier transformation, we further apply a Laplace transform to the time variable, defined as
\begin{equation}
\widetilde{\widehat{N}}(\boldsymbol{k}, E, s) \equiv \mathcal{L}_t\!\left\{ \widehat{N}(\boldsymbol{k}, E, t) \right\} = \int_0^\infty e^{-st}\widehat{N}(\boldsymbol{k}, E, t)\,dt ,
\end{equation}
From the Laplace transform property of time derivatives, we obtain
\begin{equation}
\mathcal{L}_t \left\{ \frac{\partial \widehat{N}}{\partial t} \right\} = s \widetilde{\widehat{N}}(\boldsymbol{k}, E, s) - \widehat{N}(\boldsymbol{k}, E, 0) ,
\end{equation}
where $\widehat{N}(\boldsymbol{k}, E, 0) = 0$. Applying the temporal Laplace transform to Eq.~\eqref{eq:10} yields
\begin{equation}\label{eq:14}
\begin{split}
\widetilde{\widehat{N}}(\boldsymbol{k}, E, s) ={} & \frac{1}{s + \nu(E) + D_\parallel(E)k_\parallel^2 + D_\perp(E)k_\perp^2} \\
& \times \left\{ \widetilde{\widehat{Q}}(\boldsymbol{k}, E, s) + \frac{\partial}{\partial E} \left[ b(E)\widetilde{\widehat{N}}(\boldsymbol{k}, E, s) \right] \right. \\
& \quad \left. + \int_E^{E_{\max}} T(E, E') \widetilde{\widehat{N}}(\boldsymbol{k}, E', s)\,dE' \right\} .
\end{split}
\end{equation}
To numerically solve the energy integral equation in the Fourier--Laplace space, the finite energy interval ($E_{\min} \leq E \leq E_{\max}$) is represented by $n$ logarithmically spaced energy grid points from low to high energies. The integral is discretized using trapezoidal quadrature weights associated with the energy grid, yielding
\begin{equation}\label{eq:15}
\int_{E_i}^{E_{\max}} T(E_i, E') \widetilde{\widehat{N}}(\boldsymbol{k}, E', s)\,dE' \approx \sum_{j=i+1}^{n-1} T(E_i, E_j)\, \omega_j\, \widetilde{\widehat{N}}_j ,
\end{equation}
Here, $\omega_j$ represents the quadrature weights, given by
\begin{equation}
\omega_j =
\begin{cases}
\frac{E_1 - E_0}{2} , & j = 0 , \\[6pt]
\frac{E_{j+1} - E_{j-1}}{2} , & 1 \leq j \leq n-2 , \\[6pt]
\frac{E_{n-1} - E_{n-2}}{2} , & j = n-1 .
\end{cases}
\end{equation}

The aforementioned weights can also be interpreted as the effective energy bin widths associated with the nodes $E_j$. In the continuous energy description, $\nu_j = \nu(E_j)$ represents the total physical ICS rate for electrons with energy $E_j$. However, on a finite energy grid, after a fraction of scattering events, the final energy of the electrons remains within the original energy bin. Therefore, what genuinely enters the loss term of the discretized equation should be the effective scattering rate for leaving the current energy bin, given by
\begin{equation}\label{eq:17}
\nu_j^{\rm exit} \approx \sum_{i=0}^{j-1} \omega_i T(E_i, E_j) + \nu_j^{\rm low} ,
\end{equation}
where $\nu_j^{\rm low} = \int_{m_e c^2}^{E_{\min}} T(E, E_j)\,dE$ is the scattering rate for electrons falling below the energy calculation region after scattering. Here the omission of the same-bin terms does not imply neglecting same-bin scattering, but rather arises from the exact cancellation between gains and losses.

For the synchrotron radiation term, a first-order upwind discretization is adopted, yielding
\begin{equation}\label{eq:18}
{
 \displaystyle\frac{\partial}{\partial E} \left[ b(E_i)\widetilde{\widehat{N}}(\boldsymbol{k}, E_i, s) \right]  = \frac{ b(E_{i+\frac{1}{2}}) \widetilde{\widehat{N}}_{i+1} - b(E_{i-\frac{1}{2}}) \widetilde{\widehat{N}}_i }{ \Delta E_i } ,
}
\end{equation}
where $\Delta E_i = E_{i+\frac{1}{2}} - E_{i-\frac{1}{2}}= w_i$.

Substituting Eq.~\eqref{eq:15}, \eqref{eq:17}, and \eqref{eq:18} back into Eq.~\eqref{eq:14} yields
\begin{equation}\label{eq:19}
{
\widetilde{\widehat{N}}_i = \frac{ \widetilde{\widehat{Q}}_i + \displaystyle\sum_{j=i+1}^{n-1} T(E_i, E_j)\Delta E_j \widetilde{\widehat{N}}_j + \displaystyle\frac{b(E_{i+\frac{1}{2}})}{\Delta E_i} \widetilde{\widehat{N}}_{i+1} }{ s + \nu_i^{exit} + D_\parallel(E_i)k_\parallel^2 + D_\perp(E_i)k_\perp^2 + \displaystyle\frac{b(E_{i-\frac{1}{2}})}{\Delta E_i} } .
}
\end{equation}

To simplify the anisotropic diffusion term, since the parallel and perpendicular diffusion coefficients share the same energy dependence \citep{2008ApJ...673..942Y}, we introduce the scaled Fourier variable $\mathbf{q} = \left( \sqrt{D_{0,\parallel}}\,k_\parallel, \sqrt{D_{0,\perp}}\,\mathbf{k}_\perp \right)$, yielding $q^2 = D_{0,\parallel}k_\parallel^2 + D_{0,\perp}k_\perp^2$; thus, the diffusion term can be written as $D_\parallel(E)k_\parallel^2 + D_\perp(E)k_\perp^2 = q^2 \left(\frac{E}{E_{\mathrm{ref}}}\right)^\delta$. We further define the energy vector
\begin{equation}
    \widetilde{\widehat{\mathbf{N}}}(q,s) = 
    \begin{pmatrix}
        \widetilde{\widehat{N}}_0(q,s) \\
        \widetilde{\widehat{N}}_1(q,s) \\
        \vdots \\
        \widetilde{\widehat{N}}_{n-1}(q,s)
    \end{pmatrix},
\end{equation}
Then, Eq.~\eqref{eq:19} can be written in compact form as
\begin{equation}\label{eq:21}
    \left[ s\mathbf{I} + \mathbf{A} + q^2\mathbf{F} \right] \widetilde{\widehat{\mathbf{N}}}(q,s) = \mathbf{Q}(s),
\end{equation}
where $\mathbf{I}$ is the $n \times n$ identity matrix, and the matrix $\mathbf{A}$ describes energy transitions and removal from the current energy bin induced by ICS and synchrotron radiation, written as
\begin{equation}
\begin{aligned}
    A_{ii} &= \nu_i^{\mathrm{exit}} + \displaystyle\frac{b(E_{i-\frac{1}{2}})}{\Delta E_i}, \\
    A_{ij} &= -T(E_i, E_j)\Delta E_j - \displaystyle\frac{b(E_{i+\frac{1}{2}})}{\Delta E_i} \delta_{j,i+1}, \qquad j > i, \\
    A_{ij} &= 0, \qquad j < i.
\end{aligned}
\end{equation}
Here, $\mathbf{F}$ describes the energy dependence of the diffusion coefficient,
\begin{equation}
    \mathbf{F} = \operatorname{diag}(f_0, f_1, \dots, f_{n-1}), \qquad f_i = \left(\frac{E_i}{E_{\mathrm{ref}}}\right)^\delta.
\end{equation}
The right-hand side term $\mathbf{Q}(s)$ is the energy vector of the injection term after the corresponding transformation.

When the matrix $\left[ s\mathbf{I} + \mathbf{A} + q^2\mathbf{F} \right]$ is invertible, Eq.~\eqref{eq:21} yields a unique solution, given directly by
\begin{equation}
    \widetilde{\widehat{\mathbf{N}}}(q,s) = \left[ s\mathbf{I} + \mathbf{A} + q^2\mathbf{F} \right]^{-1} \mathbf{Q}(s).
\end{equation}

Since $\mathbf{F}$ is a positive diagonal matrix, both $\mathbf{F}^{1/2}$ and $\mathbf{F}^{-1/2}$ exist and are also diagonal matrices. Defining $\mathbf{B}(s) = \mathbf{F}^{-1/2} (s\mathbf{I} + \mathbf{A}) \mathbf{F}^{-1/2},$
we obtain
\begin{equation}
    \widetilde{\widehat{\mathbf{N}}}(q,s) = \mathbf{F}^{-1/2} \left[ q^2\mathbf{I} + \mathbf{B}(s) \right]^{-1} \mathbf{F}^{-1/2} \mathbf{Q}(s).
\end{equation}

Applying the inverse Fourier transform with respect to the spatial variables to the solution $\widetilde{\widehat{\mathbf{N}}}(q,s)$ in Fourier--Laplace space, we obtain
\begin{equation}
\widetilde{{\mathbf{N}}}(\mathbf{r},s) = \frac{1}{(2\pi)^3} \int_{\mathbb{R}^3} \widetilde{\widehat{\mathbf{N}}}(q,s) e^{\mathrm{i}\boldsymbol{k}\cdot\boldsymbol{r}} \,\mathrm{d}^3 k ,
\end{equation}
To simplify the anisotropic diffusion term, we define an effective distance vector $\mathbf{R} = \left( \frac{r_\parallel}{\sqrt{D_{0,\parallel}}}, \frac{\mathbf{r}_\perp}{\sqrt{D_{0,\perp}}} \right)$, whose magnitude satisfies $R^2 = \frac{r_\parallel^2}{D_{0,\parallel}} + \frac{r_\perp^2}{D_{0,\perp}}$. Simultaneously, we have $\mathbf{k}\cdot\mathbf{r} = \mathbf{q}\cdot\mathbf{R}$ and $\mathrm{d}^3 k = \frac{\mathrm{d}^3 q}{D_{0,\perp}\sqrt{D_{0,\parallel}}}$; thus, the inverse Fourier transform can be written as
\begin{equation}
\widetilde{{\mathbf{N}}}(\mathbf{r},s) = \frac{1}{(2\pi)^3 D_{0,\perp}\sqrt{D_{0,\parallel}}} \int_{\mathbb{R}^3} \widetilde{\widehat{\mathbf{N}}}(q,s) e^{\mathrm{i}\mathbf{q}\cdot\mathbf{R}} \,\mathrm{d}^3 q .
\end{equation}

Furthermore, adopting a spherical coordinate system with the polar axis aligned along $\mathbf{R}$, we have $\mathbf{q} \cdot \mathbf{R} = q R \cos\theta$ and $\mathrm{d}^3 q = q^2 \sin\theta \, \mathrm{d}q \, \mathrm{d}\theta \, \mathrm{d}\varphi$. Since the injection term $Q$ is independent of the direction of $\mathbf{q}$, $\widetilde{\widehat{\mathbf{N}}}(q,s)$ depends only on the scalar $q^2$. Consequently, the three-dimensional inverse Fourier transform simplifies to a one-dimensional radial integral
\begin{equation}
{
\widetilde{{\mathbf{N}}}(\mathbf{r},s) = \frac{1}{2\pi^2 D_{0,\perp}\sqrt{D_{0,\parallel}}\,R} \int_0^\infty q\sin(qR)\, \widetilde{\widehat{\mathbf{N}}}(q,s) \,\mathrm{d}q .
}
\end{equation}

Since $\mathbf{F}^{-1/2}$ and $\mathbf{Q}(s)$ do not depend on the integration variable $q$, they can be moved outside the integral by linearity, yielding
\begin{equation}\label{eq:29}
\begin{split}
{}&\widetilde{\mathbf{N}}(\mathbf{r},s) = \frac{1}{2\pi^2 D_{0,\perp}\sqrt{D_{0,\parallel}}\,R} \times\\ 
&\mathbf{F}^{-1/2} \left[ \int_0^\infty q \sin(qR) \left( q^2\mathbf{I} + \mathbf{B}(s) \right)^{-1} \,\mathrm{d}q \right] \mathbf{F}^{-1/2} \mathbf{Q}(s).
\end{split}
\end{equation}

In general, a matrix function can be defined via the Cauchy integral formula as $f(\mathbf{B}) = \frac{1}{2\pi \mathrm{i}} \oint_\Gamma f(z) (z\mathbf{I} - \mathbf{B})^{-1} \,\mathrm{d}z$. Setting $f_q(z) = \frac{1}{q^2 + z}$, the integral in Eq.~\eqref{eq:29} can be written as
\begin{equation}
\begin{aligned}
{}&\int_0^\infty q \sin(qR) \left( q^2\mathbf{I} + \mathbf{B} \right)^{-1} \,\mathrm{d}q\\
&= \frac{1}{2\pi \mathrm{i}} \oint_\Gamma \left[ \int_0^\infty \frac{q \sin(qR)}{q^2 + z} \,\mathrm{d}q \right] (z\mathbf{I} - \mathbf{B})^{-1} \,\mathrm{d}z \\
&= \frac{1}{2\pi \mathrm{i}} \oint_\Gamma \frac{\pi}{2} \mathrm{e}^{-R\sqrt{z}} (z\mathbf{I} - \mathbf{B})^{-1} \,\mathrm{d}z \\
&= \frac{\pi}{2} \mathrm{e}^{-R\sqrt{\mathbf{B}}},
\end{aligned}
\end{equation}
Thus, the solution to the equation in Laplace space can be expressed as
\begin{equation}\label{eq:31}
    \widetilde{\mathbf{N}}(\mathbf{r},s) = \frac{1}{4\pi D_{0,\perp}\sqrt{D_{0,\parallel}}\,R} \mathbf{F}^{-1/2} \mathrm{e}^{-R\sqrt{\mathbf{B}}} \mathbf{F}^{-1/2} \mathbf{Q}(s),
\end{equation}
where, under the conditions that $R > 0$ and $\sigma[\mathbf{B}(s)] \cap (-\infty, 0] = \varnothing$, $\sqrt{\mathbf{B}(s)}$ denotes the principal matrix square root. Here, $\sigma[\mathbf{B}(s)]$ denotes the spectrum of the matrix $\mathbf{B}(s)$. The principal matrix square root is computed using an algorithm based on Schur decomposition \citep{Deadman2013}, while the matrix exponential is evaluated via the scaling and squaring method combined with Padé approximation \citep{HighamAlMohy2010}.

After obtaining the solution $\widetilde{{N}}(\mathbf{r},s)$ in Laplace space, its time dependence is recovered by applying the inverse Laplace transform, which is defined as
\begin{equation}
{N}(\boldsymbol{r},t) = \frac{1}{2\pi i} \int_{\gamma-i\infty}^{\gamma+i\infty} e^{st} \widetilde{{N}}(\boldsymbol{r},s)\,ds ,
\end{equation}
where $\gamma$ is the real part of the Bromwich integration line. Since the Bromwich integral is evaluated along an infinitely long vertical path in the complex plane, direct numerical integration is highly inefficient. In this work, the fixed Talbot contour is adopted to numerically evaluate the Bromwich inverse Laplace integral. According to the Talbot method, the original integration path is deformed into a contour located in the left half of the complex plane, and the contour integral is discretized via quadrature using a finite number of complex nodes and their corresponding weights \citep{10.1093/imamat/23.1.97, https://doi.org/10.1002/nme.995}.

Following the fixed Talbot discretization scheme presented by Abate and Valkó \citep{https://doi.org/10.1002/nme.995}, we obtain
\begin{equation}
{N}(\boldsymbol{r},t) \approx Re[ \sum_{m=0}^{N_T-1} c_m(t)\, \widetilde{{N}} \bigl(\boldsymbol{r}, s_m(t)\bigr)] ,
\end{equation}
where $N_T$ is the number of Talbot quadrature nodes, and we take $N_T = 28$. Here, $s_m(t)$ denotes the $m$-th sampling node along the complex integration contour, $c_m(t)$ represents the corresponding complex quadrature weight, and $\operatorname{Re}$ denotes the real part. Defining $a = \frac{2N_T}{5}$, $\theta_m = \frac{m\pi}{N_T}$, and $m = 1, 2, \dots, N_T-1$, the complex sampling nodes $s_m(t)$ are given by
\begin{equation}
\begin{split}
{}&s_0(t) = \frac{a}{t} , \qquad  s_m(t) = \frac{a\theta_m}{t} \left(\cot\theta_m + \mathrm{i}\right) , \\
& m = 1, 2, \dots, N_T-1 .
\end{split}
\end{equation}
The corresponding complex quadrature weights are expressed as
\begin{equation}
\begin{split}
{}&c_0(t) = \frac{e^a}{5t} ,\\
&c_m(t) = \frac{2}{5t} e^{a\theta_m(\cot\theta_m + \mathrm{i})} \times \\
& \quad \quad \quad \left[ 1 + \mathrm{i}\theta_m\left(1 + \cot^2\theta_m\right) - \mathrm{i}\cot\theta_m \right] ,\\
& m = 1, 2, \dots, N_T-1 .
\end{split}
\end{equation}

The numerical solution procedure involves evaluating the matrix function at each complex Talbot node, followed by performing the weighted summation specified by the inverse Laplace transform to obtain $N$.

\section{Numerical Validation and Application to the TeV Halo of Geminga Pulsar}

To verify the reliability of this numerical solving method, we set $q=0$ to obtain the spatially integrated solution and compare it with Monte Carlo simulation results \citep{g41m-jzfj} based on the same ICS Eq.~\eqref{eq:ics_kernel}. The photon fields we adopt include the CMB, the infrared radiation from dust, and the starlight radiation, with their temperatures and energy densities being ($2.7\,\mathrm{K}$, $0.26\,\mathrm{eV\,cm^{-3}}$), ($20\,\mathrm{K}$, $0.3\,\mathrm{eV\,cm^{-3}}$), and ($5000\,\mathrm{K}$, $0.3\,\mathrm{eV\,cm^{-3}}$), respectively~\citep{doi:10.1126/science.aan4880}. For the dust and starlight background, we assume that $n(\varepsilon)$ follows a graybody spectrum; that is, it adopts the spectral shape of blackbody radiation but is renormalized according to its energy density. To facilitate a direct comparison between the results obtained from different computational methods, we adopt a monoenergetic initial injection of electrons with an energy of $100\,\mathrm{TeV}$, whose injection spectrum is given by $Q(\mathbf{r}, E, t) = \delta^{(3)}(\mathbf{r}) \, \delta(t) \delta\!\left(E - 100\,\mathrm{TeV}\right)$; meanwhile, the electron energy distribution obtained from the Monte Carlo simulation is normalized to yield the corresponding electron energy probability density. Meanwhile, we set the synchrotron energy loss rate to $b(E)=0$, thereby temporarily neglecting the synchrotron radiation process.

\begin{figure*}[t]
    \centering
    \includegraphics[trim={0 0.cm 0 0}, clip, width=0.8\textwidth]{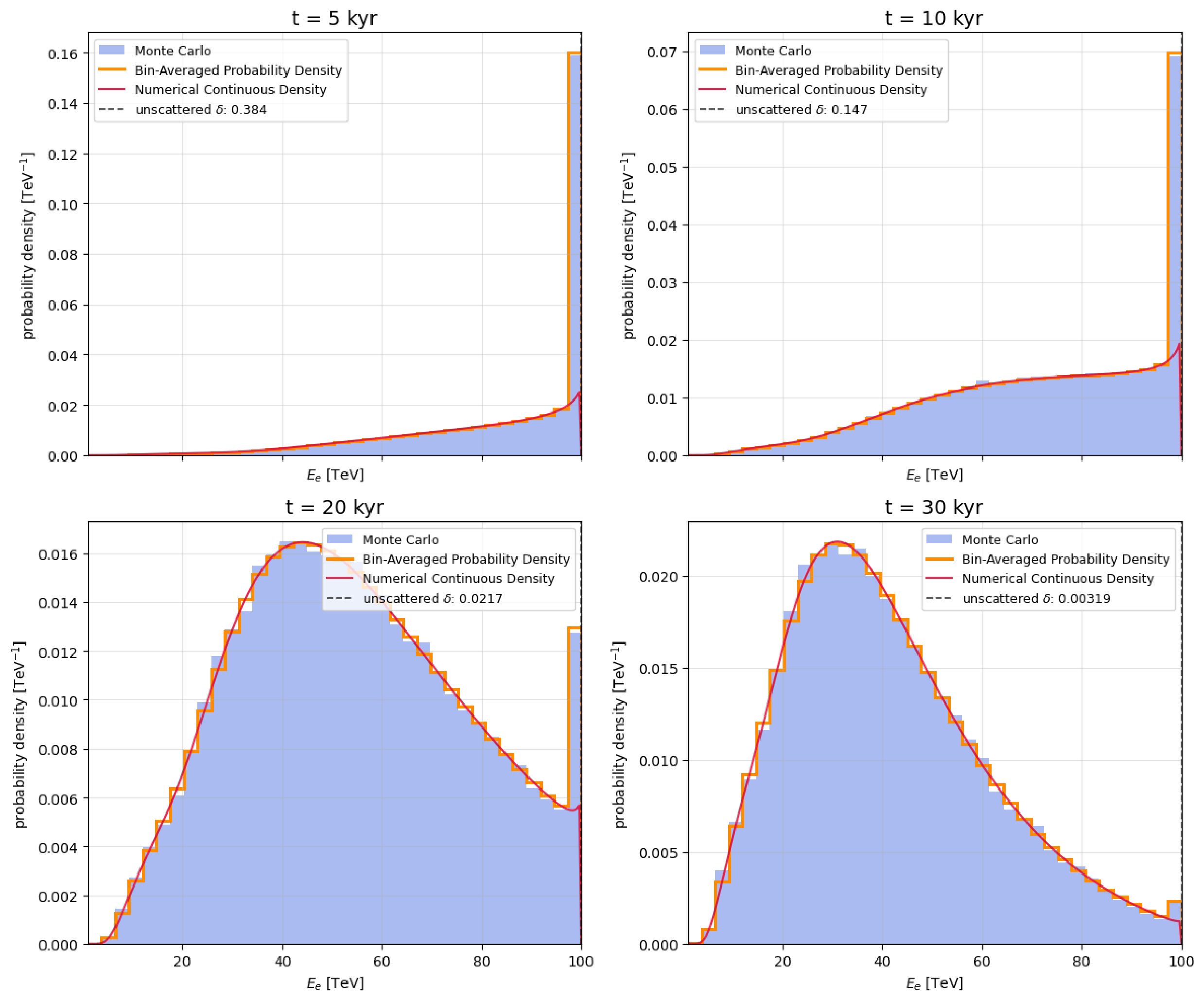}
    \caption{Comparison of Monte Carlo simulations and numerical solutions for the discrete-ICS cooling probability distribution. The four panels show the probability density distributions of electron energy $E$ at $t = 5$, $10$, $20$, and $30\,\mathrm{kyr}$, respectively. The blue histograms represent the Monte Carlo simulation results, while the red solid lines denote the continuous component of the numerical solutions, excluding the unscattered $\delta$-function component; the probabilities of electrons that have not undergone ICS are given by $\delta$ labeled in each panel. The orange stepped lines show the binned representation of the numerical probability distributions.
    }
    \label{fig:1}
\end{figure*}

Since the initial injection spectrum is a $\delta$-function, the unscattered electrons can be viewed as having infinite height and zero width at the initial energy, while possessing a finite integrated probability. Therefore, the red solid line represents only the continuous component formed by electrons that have undergone at least one ICS. To facilitate a direct comparison with the Monte Carlo histograms, we adopt the same energy binning as used in the Monte Carlo simulation to uniformly process both the continuous component and the unscattered $\delta$-component, thereby yielding the binned average probability density indicated by the orange stepped line. As shown in Figure~\ref{fig:1}, the numerical solution agrees well with the Monte Carlo simulation results, demonstrating that the numerical scheme employed in this study can accurately describe the evolution of the electron energy distribution induced by discrete ICS cooling.

%\section{Effects of Discrete Scattering}

In high-energy astrophysical sources, {PWNe can accelerate electrons to energies exceeding $100\,\mathrm{TeV}$ over extended timescales} \citep{annurev:/content/journals/10.1146/annurev.astro.44.051905.092528, 2017SSRv..207..175R}. Therefore, PWNe and pulsar halos formed by their escaping electrons \citep{10.3389/fspas.2022.1022100, doi:10.1142/S0217751X22300113} provide ideal physical scenarios for applying discrete corrections to ICS. 

The impact of discrete scattering on the evolution of the electron energy spectrum has been discussed by \citet{g41m-jzfj}. For the same injection spectrum, the continuous cooling approximation predicts evolved spectra with a lower high-energy cutoff; in contrast, discrete scattering corrections preserve a more pronounced high-energy tail. Furthermore, under the combined effects of a time-dependent injection rate and stochastic energy losses, sharp spectral features formed during evolution are smoothed out by the energy broadening induced by discrete scattering \citep{PhysRevD.107.103021}. Our findings are in good agreement with the conclusions of Refs.~\citep{PhysRevD.107.103021, g41m-jzfj}. 

Discrete scattering not only modifies the evolution of the electron energy spectrum but also influences its spatial propagation process. Compared with the continuous cooling approximation, discrete scattering both preserves high-energy surviving particles and can cause particles to skip specific electron energy regimes via single large energy losses, thereby affecting their effective propagation scale. Consequently, the direction of change in the fitted diffusion coefficient depends on the injection spectrum, magnetic field, and photon energy range. Therefore, when reproducing the spatial scale of the observed $\gamma$-ray halo, the diffusion coefficient required by the discrete scattering model differs from the value inferred from the continuous cooling model. In Ref.~\citep{doi:10.1126/science.aan4880}, the best-fit diffusion coefficient obtained by fitting the surface brightness profile is $D_{100} = 3.2 \times 10^{27} \,\mathrm{cm}^2\,\mathrm{s}^{-1}$. By setting $M_{\rm A} = 1$ while keeping all other parameters fixed, we performed a refit on the same basis. The surface brightness is calculated using the \texttt{naima} package~\citep{Khangulyan_2014, naima}. For the treatment of continuous energy loss in ICS, we adopt the following formula
\begin{equation}
\frac{dE}{dt} = -\int \int (E_\gamma - \varepsilon) f(E, E_\gamma, \varepsilon)\,dE_\gamma\,d\varepsilon.
\end{equation}

\begin{figure}[!ht]
    \centering
    \includegraphics[trim={0 0.cm 0 0}, clip, width=0.4\textwidth]{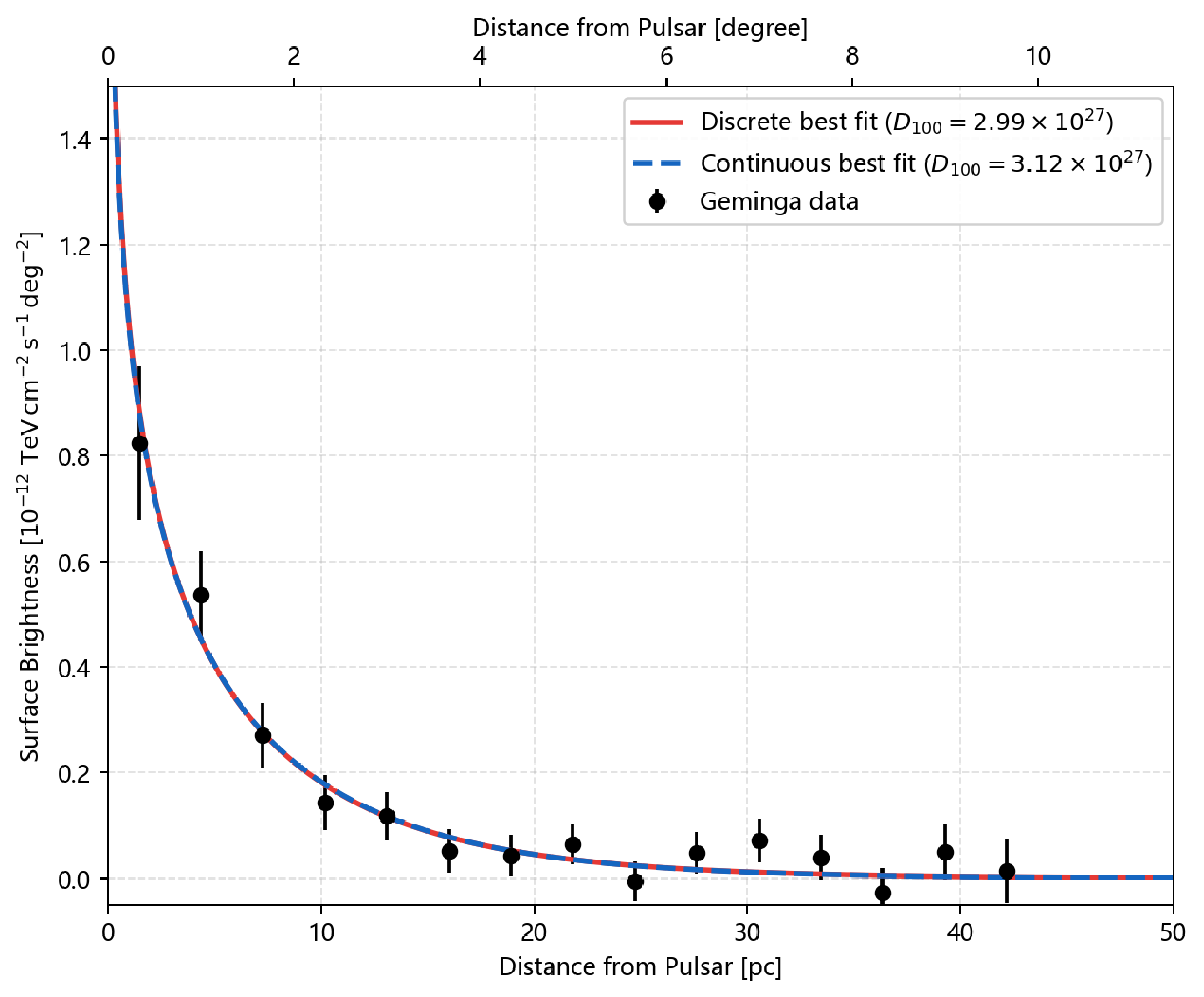}
    \caption{Surface brightness profile as a function of projected distance from the Geminga pulsar. The red curve represents the best fit for discrete ICS with $D_{100} = 2.99 \times 10^{27} \, \mathrm{cm}^2 \, \mathrm{s}^{-1}$. The blue dashed curve represents the best fit for continuous ICS with $D_{100} = 3.12 \times 10^{27} \, \mathrm{cm}^2 \, \mathrm{s}^{-1}$.
    }
    \label{fig:2}
\end{figure}
%\FloatBarrier
The best-fit diffusion coefficient for discrete ICS is slightly lower than that for continuous ICS, but the overall variation remains limited (Fig. \ref{fig:2}). This can likely be attributed to the physical parameters adopted: the injection spectral index ($\alpha = 2.24$) is relatively soft, resulting in a lower fraction of high-energy electrons; meanwhile, with the magnetic field strength set to $B = 3\,\mu\mathrm{G}$, synchrotron losses exert a non-negligible effect on the electron energy evolution, thereby partially dampening the impact of discrete ICS cooling corrections on the surface brightness profile. For harder injection spectra or lower magnetic field strengths, discrete scattering effects are expected to become more prominent, potentially leading to more pronounced changes in the best-fit diffusion coefficient.  {To examine this possibility, we adopt a cutoff power-law injection spectrum with a spectral index of $\alpha = 1.0$ and a cutoff energy of $E_{\mathrm{cut}} = 100\,\mathrm{TeV}$, and set the magnetic field strength to $B = 1\,\mu\mathrm{G}$. Meanwhile, we replace the radial surface brightness profile data used in the fit from Ref.~\citep{doi:10.1126/science.aan4880} with those from Ref.~\citep{doi:10.1126/sciadv.adv8173}. The corresponding $\gamma$-ray energy bands are $5$--$50\,\mathrm{TeV}$ and $15.8$--$251\,\mathrm{TeV}$, respectively; the latter extends to higher energies, allowing us to test the impact of discrete ICS on the propagation of high-energy electrons and the resulting spatial distribution of their emission.}

\begin{figure*}[t]
    \centering
    \includegraphics[trim={0 0.cm 0 0}, clip, width=0.8\textwidth]{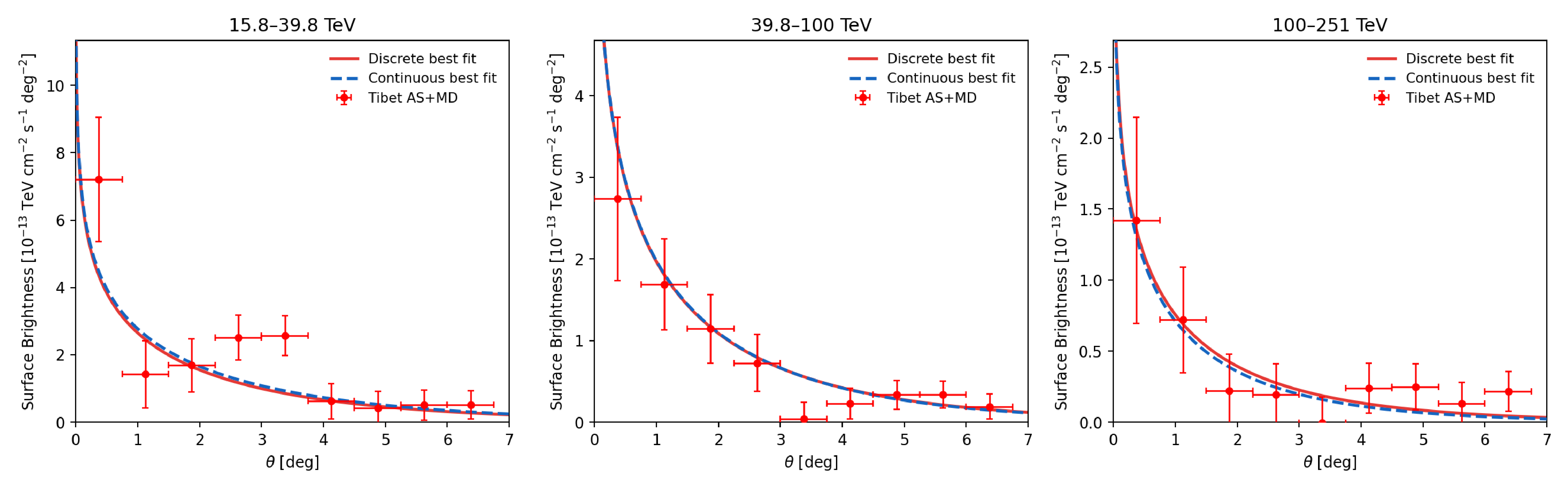}
    \caption{ {Radial surface brightness profiles of Geminga in the $15.8\text{--}39.8\text{ TeV}$, $39.8\text{--}100\text{ TeV}$, and $100\text{--}251\text{ TeV}$ energy bands. Red data points with error bars show the Tibet AS+MD observations, while the red solid lines and blue dashed lines represent the best fits of the discrete-IC and continuous-IC energy-loss models, respectively. The best-fit parameters for the discrete model are $D_{100} = 3.82 \times 10^{27}\,\mathrm{cm^2\,s^{-1}}$ and normalization $\eta = 0.0404$, whereas those for the continuous model are $D_{100} = 4.41 \times 10^{27}\,\mathrm{cm^2\,s^{-1}}$ and $\eta = 0.0463$, where $\eta$ is the fraction of the spin-down power converted into injected electron and positron power.}
    }
    \label{fig:3}
\end{figure*}

 {Under this parameter setup, the best-fit diffusion coefficient in the continuous energy-loss model is higher than that in the discrete energy-loss model by $\sim 15\%$, representing a larger relative difference between the two treatments; meanwhile, the required efficiency $\eta$ in the continuous energy-loss model is also higher by $\sim 15\%$ than in the discrete energy-loss model (Fig.\ref{fig:3}). These results indicate that under the current parameter choices and observational energy band, treating ICS discretely simultaneously affects the inferred particle diffusion efficiency and source injection power. Compared with the continuous energy-loss approximation, discrete scattering allows a fraction of electrons to retain higher energies for longer durations, thereby altering their diffusion history and their contribution to high-energy $\gamma$-ray emission. Consequently, to reproduce the same surface brightness profile, the discrete energy-loss model favors a lower diffusion coefficient and a lower energy conversion efficiency. These differences suggest that when constraining pulsar halo parameters using ultra-high-energy $\gamma$-ray observations, the discreteness of ICS can constitute a modeling effect worth taking into account.}

\section{Discussions and Conclusions}

Building upon the conventional transport equation, this paper incorporates a discrete ICS collision operator constructed from the Klein–Nishina differential scattering rate \citep{RevModPhys.42.237}, establishing a transport model capable of simultaneously describing the stochastic energy loss and spatial propagation of electrons. Compared with the continuous cooling approximation, the discrete model no longer treats electron energy evolution as a deterministic trajectory, but instead yields a time-broadened energy distribution.

We perform a conservative discretization in energy space, transform the diffusion equation into a finite-dimensional matrix equation using spatial Fourier and temporal Laplace transforms, and execute the spatial inversion over the infinite Fourier domain via matrix function methods. A numerical inverse Laplace transform is implemented using the Talbot contour integral \citep{https://doi.org/10.1002/nme.995}, thereby obtaining the joint spatial and energy distribution of electrons at any given time. This approach does not rely on Monte Carlo sampling, significantly reducing computational costs compared to traditional Monte Carlo methods. 

The reliability of the discrete collision operator and its numerical solver is validated through comparison with Monte Carlo results. The spectral evolution results align with previous conclusions regarding discrete ICS cooling \citep{g41m-jzfj}, serving primarily as a benchmark for the model and algorithm in this work. On this basis, we further calculate the spatial distribution of electrons and the ICS $\gamma$-ray surface brightness under continuous injection conditions, thereby extending the impact of discrete scattering from spectral evolution to electron diffusion and the spatial morphology of photon halos. Fitting the observed surface brightness using Geminga parameters identical to those in Ref.~\citep{doi:10.1126/science.aan4880} (except for the diffusion coefficient) yields $D_{100} = 2.99 \times 10^{27} \,\mathrm{cm}^2\,\mathrm{s}^{-1},$ indicating that discrete ICS not only affects the electron energy spectrum but also influences the diffusion coefficient inferred from surface brightness profiles by altering the residence time and effective propagation scale of high-energy electrons. Under the current parameters, the magnitude of this correction remains small, which can likely be attributed to the relatively soft injection spectrum and the dampening effect of synchrotron losses; for harder injection spectra or lower magnetic field strengths, the influence of discrete scattering on spatial propagation is expected to become more prominent.  {Indeed, when adopting a harder cutoff power-law injection spectrum ($\alpha = 1.0$, $E_{\mathrm{cut}} = 100\,\mathrm{TeV}$) and a weaker magnetic field ($B = 1\,\mu\mathrm{G}$) to fit the higher-energy Tibet AS+MD surface brightness profiles ($15.8$--$251\,\mathrm{TeV}$), the relative difference between the continuous and discrete energy-loss models increases to $\sim 15\%$ for both the best-fit diffusion coefficient and the required injection efficiency $\eta$.}

Overall, the primary contribution of this work lies in establishing a transport framework that couples discrete inverse Compton cooling with spatial diffusion, as well as proposing a deterministic solution method based on Fourier–Laplace transforms and matrix functions. This framework provides a unified computational foundation for systematically investigating the stochastic cooling, spatial propagation, and $\gamma$-ray emission characteristics of ultra-high-energy electrons.

\begin{acknowledgments}
%We thank Haiming Zhang, Ruoyu Liu, Yang Su, L.D.Anderson, P.P.Delia for invaluable discussions. This work makes use of molecular line data from the Boston University-FCRAO Galactic Ring Survey (GRS).
This work is supported by the National Natural Science Foundation of China under the grant 12375103 and the Rongpiao innovation team program of Chengdu. 
%, the Natural Science Foundation for Young Scholars of Sichuan Province, China (No. 2022NSFSC1808), and the Fundamental Research Funds for the Central Universities (No. 2682022ZTPY013).
\end{acknowledgments}

\end{document}